\documentclass[twocolumn,showpacs,superscriptaddress,amssymb,10pt]{revtex4}

\usepackage{graphicx}
\usepackage{dcolumn}
\usepackage{bm}
\usepackage{epsfig}
\usepackage{color}
\usepackage{longtable}
\usepackage{sidecap}
\usepackage{float}
\usepackage{multirow}
\usepackage{ulem}  
\usepackage{mathrsfs}
\usepackage{amsmath}
\usepackage{extarrows}

\definecolor{aogreen}{rgb}{0.0, 0.5, 0.0}

\def\ketm#1{  \left\vert  #1   \right\rangle   }

\def\sprm#1#2{  \left\langle #1 \left\vert \right. #2 \right\rangle   }

\def\redmem#1#2#3{  \left\langle #1 \left\Vert
                  #2 \right\Vert #3 \right\rangle   }
\begin{document}

\preprint{}
%
\title{Angle-resolved x-ray spectroscopic scheme to determine overlapping hyperfine splittings in highly charged helium-like ions}

\author{Z.~W.~Wu}
\affiliation{Helmholtz-Institut Jena, D-07743 Jena, Germany}
\affiliation{Key Laboratory of Atomic and Molecular Physics $\&$ Functional Materials of Gansu Province,
             College of Physics and Electronic Engineering, Northwest Normal University, Lanzhou 730070, P.R. China}

\author{A.~V.~Volotka}
\affiliation{Helmholtz-Institut Jena, D-07743 Jena, Germany}
\affiliation{Department of Physics, St.~Petersburg State University, 198504 St.~Petersburg, Russia}

\author{A.~Surzhykov}
\affiliation{Physikalisch-Technische Bundesanstalt, D-38116 Braunschweig, Germany}
\affiliation{Technische Universit\"at Braunschweig, D-38106 Braunschweig, Germany}


\author{S.~Fritzsche}
\affiliation{Helmholtz-Institut Jena, D-07743 Jena, Germany}
\affiliation{Theoretisch-Physikalisches Institut, Friedrich-Schiller-Universit\"at Jena, D-07743 Jena, Germany}

\date{\today \\[0.3cm]}

\begin{abstract}

An angle-resolved x-ray spectroscopic scheme is presented for determining the hyperfine splitting of highly charged ions. For
helium-like ions, in particular, we propose to measure either the angular distribution or polarization of the $1s2p~^{3}P_{1}, F
\rightarrow 1s^{2}~^{1}S_{0}, F_{f}$ emission following the stimulated decay of the initial $1s2s~^{1}S_{0}, F_{i}$ level. It is
found that both the angular and polarization characteristics of the emitted x-ray photons strongly depends on the (relative)
\textit{splitting} of the partially overlapping hyperfine $1s2p~^{3}P_{1}, F$ resonances and may thus help resolve their
hyperfine structure. The proposed scheme is feasible with present-day photon detectors and allows a measurement of the hyperfine
splitting of helium-like ions with a relative accuracy of about $10^{-4}$.

\end{abstract}


\maketitle

\textit{Introduction.}---Hyperfine splitting of energy levels occurs primarily due to the interaction of bound electrons with the
magnetic (dipole) field of nucleus. The strength of this nuclear magnetic field increases rapidly with the nuclear charge and
reaches about 10$^{9}$~T at the surface of $^{209}$Bi nucleus, several orders of magnitude higher than the field of the most
powerful magnets. For this reason, the study of the hyperfine splitting in highly charged ions has attracted much recent
attention from both theory and experiment, and aims to probe bound-state QED at extreme electric and magnetic fields. In the
past, various high-precision measurements on the hyperfine splitting in hydrogen-like ions were performed \cite{Klaft/PRL:1994,
Lopez-Urrutia/PRL:1996-1998, Seelig/PRL:1998, Beiersdorfer/PRA:2001} and thus stimulated a good deal of theoretical developments,
see review \cite{Volotka/AP:2013} and references therein. Due to recent advances in experiment, moreover, the accuracy of the
(measured) hyperfine splitting in $^{209}$Bi$^{82+}$ was improved by almost one order of magnitude and meanwhile reached the
level of about 10$^{-5}$ \cite{Ullmann/JPB:2015}. Until now, however, further theoretical progress has been restricted by the
lack of knowledge of the nuclear magnetization distribution. In Ref.~\cite{Shabaev/PRL:2001}, it was therefore proposed to
consider the specific difference of the hyperfine splitting in different electronic configurations, e.g., the difference between
the ground-state hyperfine splitting in hydrogen- and lithium-like ions of the same isotope, for which the uncertainty due to the
nuclear magnetization distribution is substantially reduced.

Besides ground-state hyperfine splitting in hydrogen-like ions, until now only a very few measurements have been made for
lithium- and beryllium-like praseodymium ions \cite{Beiersdorfer/PRL:2014} and lithium-like bismuth ions
\cite{Lochmann/PRA:2014}. Recently, the LiBELLE collaboration presented a new high-precision measurement of the ground-state
hyperfine splitting in $^{209}$Bi$^{80+}$ ion \cite{Ullmann/NC:2017}. The specific difference between the hyperfine splitting in
hydrogen- and lithium-like bismuth ions, determined in this measurement, yields 7$\sigma$ disagreement when compared with the
corresponding theoretical values \cite{Volotka/PRL:2012}. In order to resolve this discrepancy, additional measurements of the
hyperfine splitting with different electronic configurations are highly desirable.

Helium-like ions are another alternative that may serve for the same purpose. Apart from the hyperfine quenching (cf.~the review
by Johnson \cite{Johnson/CJP:2011}), however, to the best of our knowledge there are no experimental studies on the hyperfine
structure of such ions. In contrast to hydrogen- and lithium-like ions, there is no hyperfine splitting in the $
1s^{2}~^{1}S_{0}, F_{f}$ ground level of helium-like ions. As for the excited levels $1s2p~^{1,3}P_{1}, F$ (and $1s2s~^{3}S_{1},
F'$), for which the natural linewidth is comparable in magnitude or even larger than the corresponding hyperfine splitting, they
can therefore not be resolved by conventional fluorescence spectroscopy. For the $1s2p~^{3}P_{1}, F$ levels, for example, the
natural linewidth goes rapidly from 0.10~eV for $^{71}$Ga$^{29+}$ to 12.75~eV for $^{209}$Bi$^{81+}$, while the hyperfine
splitting just increases from 0.11~eV to 5.35~eV. Moreover, the transition energies of the (partially) overlapping
$1s2p~^{3}P_{1}, F$ levels to the ground state are quite large and thus less suitable for precision measurements.

\begin{figure}[!tbp]
\includegraphics[width=0.98\linewidth]{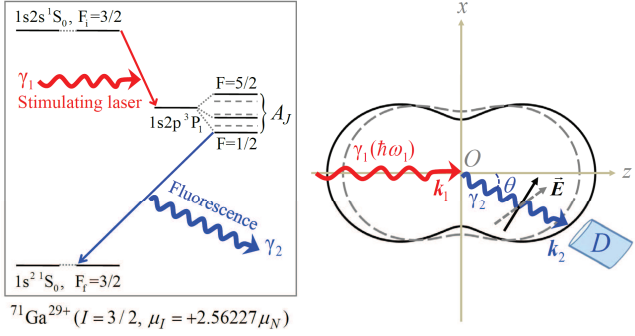}
\caption{\label{Fig-schematic} (Color online) Schema (left panel) and geometry (right panel) for measuring the two-step radiative
decay (\ref{particular-process}) of helium-like $^{71}$Ga$^{29+}$ ions. While the first-step decay is stimulated by laser photons
$\gamma_{1}$ with energy $\hbar\omega_{1}$, the fast subsequent spontaneous decay to the $ 1s^{2}~^{1}S_{0}, F_{f}$ ground level
gives rise to an emission of the $\gamma_{2}$ fluorescence photons. As distinguished in black solid and gray dashed lines, the
angular distribution and polarization of the $\gamma_{2}$ photons depend sensitively on the hyperfine constant of the
intermediate levels.}
\end{figure}

In this contribution, we propose a novel scheme for resolving the hyperfine splitting in highly charged helium-like ions by
measuring the angular distribution and angle-resolved polarization of the emitted fluorescence photons. As an example, we shall
predict and analyze the angular distribution and linear polarization of the $\gamma_{2}$ photons emitted in the two-step
radiative decay
\begin{eqnarray}
\label{particular-process}
   1s2s ~^{1}S_{0}, F_{i} + \gamma_{1}
   & \longrightarrow &  1s2p~^{3}P_{1}, F
   \nonumber \\ [0.1cm]
   & \longrightarrow &  1s^{2}~^{1}S_{0}, F_{f} + \gamma_{2}
\end{eqnarray}
of helium-like ions, from which we aim to determine the hyperfine splitting of the intermediate $1s2p~^{3}P_{1}, F$ levels.
Fig.~\ref{Fig-schematic} displays the (involved) fine-structure levels of helium-like $^{71}$Ga$^{29+}$ ions and their hyperfine
splittings (left panel). In this scheme, the ions first decay from the initial $1s2s~^{1}S_{0}, F_{i}$ level to the intermediate
$1s2p~^{3}P_{1}, F$ levels under a stimulation by incident laser photons $\gamma_1$. Subsequently, fast spontaneous decay of the
$1s2p~^{3}P_{1}, F$ levels into the ground level $1s^{2}~^{1}S_{0}, F_{f}$ occurs with an emission of the $\gamma_{2}$ photons.
The angular distribution and polarization of the emitted $\gamma_{2}$ photons are then measured as functions of the photon energy
$\hbar\omega_{1}$. The obtained $\gamma_{2}$ angular distribution and polarization are expected to be sensitive to the
$1s2p~^{3}P_{1}, F$ hyperfine splittings due to their different populations following the first-step stimulated decay, as shown
in the right panel of Fig.~\ref{Fig-schematic}.

It is well known, both experimentally \cite{Henderson/PRL:1990} and theoretically \cite{Surzhykov/PRA:2013,Wu/PRA-89:2014}, that
the angular distribution and polarization of $K\alpha_1$ photons in helium-like ions are often affected by the hyperfine
admixture of different fine-structure levels, which may alter the branching ratios between different hyperfine sublevels. In the
proposed scheme, the hyperfine levels can be populated in a controlled manner by just tuning the photon energy $\hbar\omega_{1}$
of the incident laser over the resonance of the first-step decay. By analyzing the angular distribution or linear polarization of
the emitted $\gamma_{2}$ fluorescence photons, one can restore the created population. Below, we will demonstrate that the
angular distribution and polarization of the $\gamma_{2}$ photons depend sensitively on the hyperfine splitting, as shown in the
right panel of Fig.~\ref{Fig-schematic}. The obtained results suggest that an angle- or polarization-resolved measurement of the
$1s2p~^{3}P_{1}, F$ $\rightarrow$ $1s^{2}~^{1}S_{0}, F_{f}$ emission line may provide an experimental determination on the
splitting of the partially overlapping hyperfine levels with a relative accuracy of about $10^{-4}$.

\begin{table*}[htbp]
\caption{\label{Table-1} Table of isotopes with nuclear spin $I$ and magnetic moment $\mu_{I}$ in units of the nuclear magneton
$\mu_{N}$ considered in this work \cite{Stone/ADNDT:2005}. The calculated hyperfine constant $A_{J}$ (eV) and natural linewidth
$\Gamma_{F}$ (eV) of the $1s2p~^{3}P_{1}$ level are presented, together with the transition energies (eV) of both the
$1s2s~^{1}S_{0}$ $\rightarrow$ $1s2p~^{3}P_{1}$ and $1s2p~^{3}P_{1}$ $\rightarrow$ $1s^{2}~^{1}S_{0}$ lines taken from
Ref.~\cite{Artemyev/PRA:2005}. The anisotropy and polarization sensitivity coefficients $\frac{d\beta}
{dA_{J}}\frac{A_{J}}{\beta}$ and $\frac{dP_{1}}{dA_{J}}\frac{A_{J}}{P_{1}}$ are calculated at the resonance energies.}
\begin{ruledtabular}
\begin{tabular}
{ccccccccc} Isotope        & Spin & Magnetic moment& Hyperfine constant         & Linewidth
                                  & \multicolumn{2}{c}{Transition energies}
                                  & \multirow{2}{*}{$\frac{d\beta}{dA_{J}}\frac{A_{J}}{\beta}$}
                                  & \multirow{2}{*}{$\frac{dP_{1}}{dA_{J}}\frac{A_{J}}{P_{1}}$}                             \\
           $^{A}X$         & $I$  & $\mu_{I}$      & $A_{J} \, (2~^{3}P_{1})$    & $\Gamma_{F} \, (2~^{3}P_{1})$
                                  & $2~^{1}S_{0}$ $\rightarrow$ $2~^{3}P_{1}$ & $2~^{3}P_{1}$ $\rightarrow$ $1~^{1}S_{0}$
                                  &  &                   \\
\hline
$^{\phantom{0}71}$Ga$^{29+}$ & 3/2 & +2.56227 & 0.0274 & \phantom{0}0.1012 & \phantom{0}0.173 & \phantom{0}9574.446 & $-1.184$ & $-1.356$ \\
$^{141}$Pr$^{57+}$ & 5/2 & +4.2754\phantom{0} & 0.2530 & \phantom{0}2.9551 &           12.461 &           36391.292 & $-0.400$ & $-0.499$ \\
$^{209}$Bi$^{81+}$ & 9/2 & +4.1103\phantom{0} & 0.5349 &           12.7539 &           63.961 &           76131.359 & $-0.291$ & $-0.369$ \\
\end{tabular}
\end{ruledtabular}
\end{table*}

\textit{Theoretical background.}---To understand how the splitting of overlapping hyperfine levels affect the fluorescence
emission of helium-like ions, let us start from a theoretical analysis of the photon angular distribution and linear
polarization. Our theory is developed within the framework of density matrix and second-order perturbation theory. For the
two-step decay process (\ref{particular-process}), if the geometry in Fig.~\ref{Fig-schematic} is adopted, the second-order
transition amplitudes can be expressed in the form \cite{Wu/PRA:2016}
\begin{widetext}
\begin{eqnarray}
\label{2nd-order-transition-amplitude}
  \mathcal{M}^{\lambda_{1}, \lambda_{2}}_{M_{i}, M_{f}} (\hbar\omega_{1})
         = \sum_{F M}  \sum_{p_{1} L_{1} M_{L_{1}}}  \sum_{p_{2} L_{2} M_{L_{2}}}  i^{-L_{1}-L_{2}} \,
          (i\lambda_{1})^{p_{1}} \, (i\lambda_{2})^{p_{2}} \, \delta_{\lambda_{1} \, M_{L_{1}}} \,
          D^{L_{2}}_{M_{L_{2}} \lambda_{2}} (\varphi, \theta, 0) \,
          [L_{1}, L_{2}]^{1/2} \, [F_{i}, F]^{-1/2} \, (-1)^{F_{i}-F_{f}}
          \nonumber \\ [0.1cm] \times
          \sprm{F_{f} M_{f}, L_{2} M_{L_{2}}}{F M} \, \sprm{F M, L_{1} M_{L_{1}}}{F_{i} M_{i}} \,
          \frac{ \redmem{F_{f}} {\sum_{m} \bm{\alpha}_{m} \bm{a}^{p_{2}}_{L_{2}} (\bm{r}_{m})} {F}
                 \redmem{F} {\sum_{m} \bm{\alpha}_{m} \bm{a}^{p_{1}}_{L_{1}} (\bm{r}_{m})} {F_{i}} }
               { E_{F_{i}} - E_{F} - \hbar\omega_1 + i \, \Gamma_{F} / 2 } . \, \,
\end{eqnarray}
\end{widetext}
Here, $\delta_{\lambda_{1} \, M_{L_{1}}}$ denotes a Kronecker delta function, $[a, b] \equiv (2a+1)(2b+1)$, and the standard
notations for the Wigner $D$-functions and the Clebsch-Gordan coefficients have been employed. Moreover, the individual photons
$\gamma_{1, 2}$ are characterized in terms of their helicity $\lambda$ and multipolarities $pL$, with $p=0$ for magnetic
multipoles and $p=1$ for electric ones. $E_{F}$ and $\Gamma_{F}$ represent, respectively, the energy and natural linewidth of the
hyperfine level $\ketm{F} \equiv \ketm{\alpha J I F}$ with (total) angular momentum $F$, nuclear spin $I$, angular momentum $J$
of the electronic state, and all additional quantum numbers $\alpha$ that are needed for its unique specification. It should be
noted that, we here neglect the linewidth $\Gamma_{F_{i}}$ of the initial level in the denominator as it is much smaller than the
linewidth $\Gamma_{F}$ of the intermediate levels. The hyperfine transition amplitudes $\redmem{F'} {\sum_{m} \bm{\alpha}_{m}
\bm{a}^{p}_{L} (\bm{r}_{m})} {F}$ can be obtained from the corresponding fine-structure transition amplitudes by representing the
$IJ$ coupled atomic basis states in their product basis. From the hyperfine transition amplitudes, we can easily determine the
decay rate of excited hyperfine-resolved levels and their natural linewidths. The transition amplitudes of the $1s2s~^{1}S_{0}$
$\rightarrow$ $1s2p~^{3}P_{1}$ and $1s2p~^{3}P_{1}$ $\rightarrow$ $1s^{2}~^{1}S_{0}$ are here evaluated within the framework of
perturbation theory and by including the first-order interelectronic-interaction correction, see Ref.~\cite{Indelicato/PRA:2004}
for details. For the case of the transition $1s2p~^{3}P_{1}$ $\rightarrow$ $1s^{2}~^{1}S_{0}$, the obtained results agree fairly
with previously published values \cite{Drake/PRA:1979,Johnson/AAMOP:1995,Andreev/PRA:2009}.

Apart from the hyperfine transition amplitudes and natural linewidths $\Gamma_{F}$, we still need to know the hyperfine structure
energies in order to compute the second-order transition amplitudes (\ref{2nd-order-transition-amplitude}). Since the
electric-quadrupole hyperfine interactions are negligibly small throughout the helium-like isoelectronic sequence when compared
to the nuclear magnetic-dipole interaction, they will not be considered here. The magnetic-dipole hyperfine splitting of a
fine-structure level $\alpha J$ can be expressed as
\begin{eqnarray}
\label{hf-M1-energy-shift-JJ}
   \Delta E_{\alpha JIF}^{\rm hf} = A_{J} \left[F(F+1)-I(I+1)-J(J+1)\right] / 2 \,
\end{eqnarray}
in terms of the hyperfine constant $A_{J}$. With this notation, $E_{F} = E_{\alpha J} + \Delta E_{\alpha JIF}^{\rm hf}$, where
$E_{\alpha J}$ is the energy of the corresponding fine-structure level. As seen from Eq.~(\ref{hf-M1-energy-shift-JJ}), the
hyperfine splitting can be easily obtained once the hyperfine constant is determined. The hyperfine constant of the
$1s2p~^{3}P_{1}$ level is evaluated in the intermediate coupling scheme of mixing $1s2p~^{3}P_{1}$ and $1s2p~^{1}P_{1}$ levels as
in Ref.~\cite{Volotka/CJP:2002}, while the remaining interelectronic-interaction corrections are accounted for by a local
screening potential. The effect of nuclear magnetization distribution is calculated by employing the nuclear single-particle
model \cite{Shabaev/PRA:1997}. The obtained hyperfine constants are in fair agreement with the results from
Ref.~\cite{Johnson/PRA:1997}. In addition, we also estimate the one-electron QED corrections --- self energy and vacuum
polarization --- to the hyperfine constant $A_{J}$. These corrections contribute, for instance, about 0.5\% for
$^{209}$Bi$^{81+}$ ions.

As a first attempt on this kind of studies, the incident stimulating laser is assumed to be unpolarized for simplicity. In this
case, the density matrix of the emitted $\gamma_{2}$ photons can be expressed in terms of the second-order amplitudes
(\ref{2nd-order-transition-amplitude}) as follow,
\begin{eqnarray}
\label{SimplifiedDMofEmittedPhoton}
  \rho^{\gamma_{2}}_{\lambda_{2}, \lambda_{2}^{'}}
    \equiv \langle \hat{\bm{k}}_{2}, \lambda_{2} | \rho^{\gamma_{2}} | \hat{\bm{k}}_{2}, \lambda_{2}^{'} \rangle
      \hspace{3.1cm} \nonumber \\ [0.1cm]
    = \frac{1}{2[F_{i}]} \sum_{M_{i}, M_{f}} \sum_{\lambda_{1}=\pm1}
      \mathcal{M}^{\lambda_{1}, \lambda_{2}}_{M_{i}, M_{f}} (\hbar\omega_{1})
      \mathcal{M}^{\lambda_{1}, \lambda_{2}^{' \, *}}_{M_{i}, M_{f}} (\hbar\omega_{1}) \, .
\end{eqnarray}
Once we obtain the density matrix (\ref{SimplifiedDMofEmittedPhoton}), the angular distribution and polarization of the
$\gamma_{2}$ photons can be given in terms of its matrix elements. If, for instance, the polarization of the $\gamma_{2}$ photons
remains unobserved, the $\gamma_{2}$ angular distribution follows simply from the trace of
Eq.~(\ref{SimplifiedDMofEmittedPhoton}),
\begin{eqnarray}
\label{ADofEmittedPhoton}
  \frac{d\sigma}{d\Omega}
     = \rho^{\gamma_{2}}_{+1, +1} + \rho^{\gamma_{2}}_{-1, -1}
     = \frac{\sigma_{0}}{4\pi} [1 + \beta P_{2}(\cos\theta)] \, .
\end{eqnarray}
As the $\gamma_{1}$ photons are unpolarized, the angular distribution (\ref{ADofEmittedPhoton}) is azimuthally symmetric and thus
independent of the angle $\varphi$. For this reason, it is parameterized by a single, so-called \textit{anisotropy} parameter
$\beta$ within the E1 approximation, as shown in the second equality. Moreover, the linear polarization $P_{1}$ can be given as
follows,
\begin{eqnarray}
\label{Pola.ofEmittedPhoton}
  P_{1} = (\rho^{\gamma_{2}}_{+1, -1} + \rho^{\gamma_{2}}_{-1, +1}) /
          (\rho^{\gamma_{2}}_{+1, +1} + \rho^{\gamma_{2}}_{-1, -1}) \, .
\end{eqnarray}
We are now ready to study the angular distribution and linear polarization of the emitted $\gamma_{2}$ photons.

\begin{figure}[!t]
\includegraphics[width=0.95\linewidth]{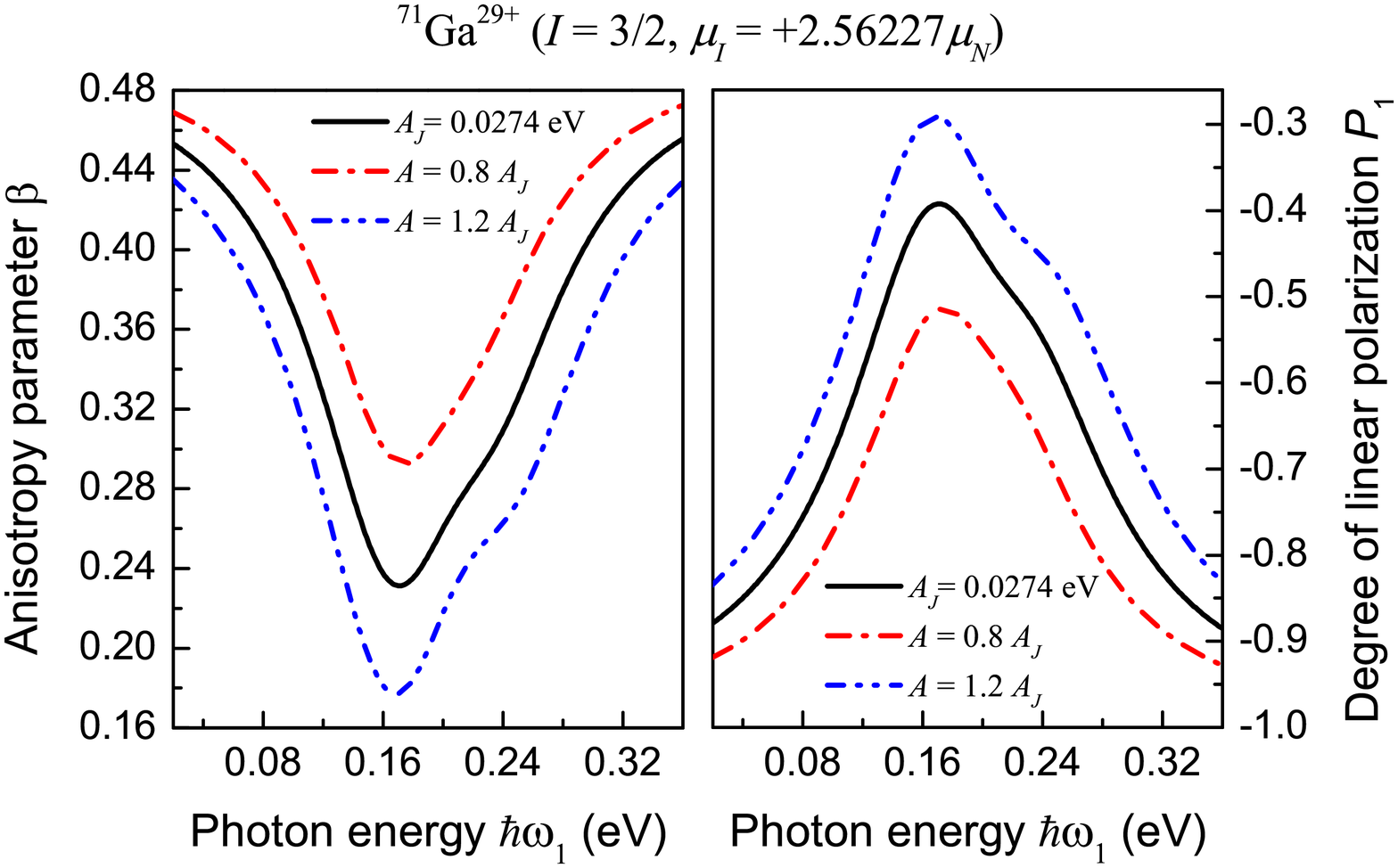}
\caption{\label{Fig-Ga71-Aniso-Polari} (Color online) Anisotropy parameter $\beta$ (left panel) and degree of linear polarization
$P_{1}$ (right panel) of the hyperfine $1s2p~^{3}P_{1}, F=1/2, 3/2, 5/2$ $\rightarrow$ $1s^{2}~^{1}S_{0}, F_{f}=3/2$ fluorescence
emission from helium-like $^{71}$Ga$^{29+}$ ions as functions of photon energy $\hbar \omega_{1}$ of the incident photons
$\gamma_{1}$. Results are shown for the calculated hyperfine constant $A_{J}=0.0274$~eV (black solid lines) as listed in
Table~\ref{Table-1} as well as for two assumed values, 0.8$A_{J}$ (red dash-dotted lines) and 1.2$A_{J}$ (blue dash-dot-dotted
lines), which differ by just 20\%.}
\end{figure}

\textit{Results and discussion.}---Table~\ref{Table-1} lists the calculated hyperfine constant and linewidth of the
$1s2p~^{3}P_{1}$ level, together with the transition energies of the $1s2s~^{1}S_{0}$ $\rightarrow$ $1s2p~^{3}P_{1}$ and
$1s2p~^{3}P_{1}$ $\rightarrow$ $1s^{2}~^{1}S_{0}$ lines from Ref.~\cite{Artemyev/PRA:2005} for heliumlike $^{71}$Ga$^{29+}$,
$^{141}$Pr$^{57+}$ and $^{209}$Bi$^{81+}$ ions. These data are used for the evaluation of the second-order transition amplitudes
and, further, to analyze the angular distribution and linear polarization of the emitted $\gamma_2$ photons.
Fig.~\ref{Fig-Ga71-Aniso-Polari} displays the anisotropy parameter $\beta$ and the linear polarization $P_{1}$ of the
$1s2p~^{3}P_{1}, F=1/2, 3/2, 5/2$ $\rightarrow$ $1s^{2}~^{1}S_{0}, F_{f}=3/2$ fluorescence photons of helium-like
$^{71}$Ga$^{29+}$ ions as functions of the photon energy $\hbar\omega_{1}$. The linear polarization $P_{1}$ is presented for the
$\gamma_{2}$ photons that are emitted perpendicular to the incident $\gamma_{1}$ photons, i.e., at $\theta = 90^{\circ}$. The
parameter $P_{1} = (I_{0^{\circ}} - I_{90^{\circ}}) / (I_{0^{\circ}} + I_{90^{\circ}})$ characterizes the intensities of the
emitted $\gamma_{2}$ photons linearly polarized in parallel ($I_{0^{\circ}}$) or perpendicular ($I_{90^{\circ}}$) to the reaction
plane defined by the propagation directions of the $\gamma_{1}$ and $\gamma_{2}$ photons. Results are shown for the calculated
hyperfine constant $A_{J}=0.0274$~eV and also for two assumed values, 0.8$A_{J}$ and 1.2$A_{J}$, which differ by just 20\%.

As can be seen from the figure, both the $\gamma_{2}$ anisotropy and linear polarization appear to be rather sensitive with
regard to the photon energy $\hbar\omega_{1}$ for any given hyperfine constant of the $1s2p~^{3}P_{1}, F$ levels. Typically, the
$\gamma_{2}$ photons have the smallest anisotropy and polarization near the resonance energy $\hbar\omega_{1} \simeq 0.173$~eV,
but both become more and more anisotropic or (linearly) polarized when the photon energy $\hbar\omega_{1}$ is tuned away from the
resonance. This dependence arises from the finite linewidths of the overlapping resonances $1s2p~^{3}P_{1}, F = 1/2, 3/2, 5/2$,
which lead to a coherent population of them during the first-step stimulated decay and, ultimately, affects the angular and
polarization behaviors of the emitted $\gamma_{2}$ photons. Moreover, both the anisotropy parameter and linear polarization
depend strongly on the hyperfine constant of the $1s2p~^{3}P_{1}$ level, especially, if the $\gamma_{1}$ photon energy is close
to the resonance, say, $\hbar\omega_{1} \simeq 0.173$~eV. The linear polarization $P_{1}$, for instance, changes from $-0.39$ for
$A_{J}=0.0274$~eV to $-0.51$ for 0.8$A_{J}$ at this resonance energy. In order to further analyze this dependence quantitatively,
two sensitivity coefficients $\frac{d\beta}{dA_{J}}\frac{A_{J}}{\beta}$ and $\frac{dP_{1}}{dA_{J}}\frac{A_{J}}{P_{1}}$ are
introduced. These coefficients reach their respective maximums at the resonance energy, which are listed in Table~\ref{Table-1}.
From these coefficients, it is quite easy to see how a change in the hyperfine constant $A_{J}$ will affect the $\gamma_{2}$
anisotropy or polarization, i.e., if $A_{J}$ is modified by, say, 20\%, $\beta$ will change by 23.6\% while $P_{1}$ by 27.2\%, as
shown in Fig.~\ref{Fig-Ga71-Aniso-Polari}.

\begin{figure}[!b]
\includegraphics[width=0.95\linewidth]{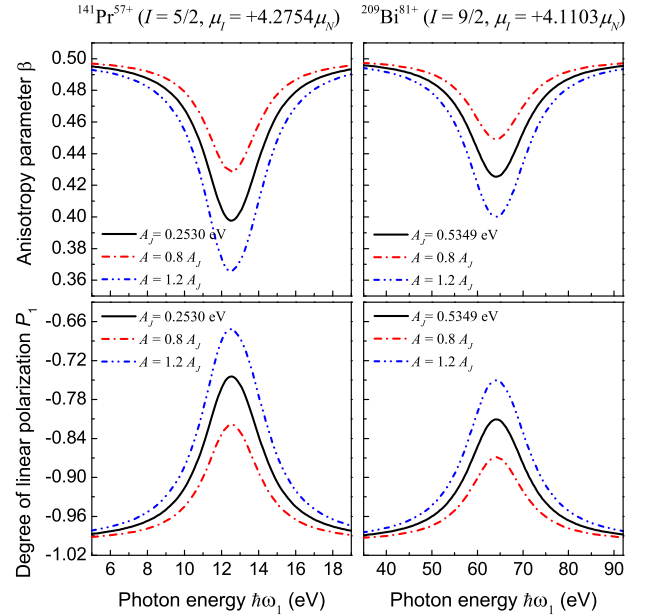}
\caption{\label{Fig-Pr141-Bi209-Aniso-Polari} (Color online) The same as in Fig.~\ref{Fig-Ga71-Aniso-Polari} but for
$^{141}$Pr$^{57+}$ (left panel) and $^{209}$Bi$^{81+}$ (right panel) ions.}
\end{figure}

Besides the low-$Z$ $^{71}$Ga$^{29+}$, we also consider medium- and high-$Z$ ions such as $^{141}$Pr$^{57+}$ and
$^{209}$Bi$^{81+}$. For these ions, the corresponding $\gamma_{2}$ anisotropy and linear polarization still depend on the
$1s2p~^{3}P_{1}, F$ hyperfine splittings and also on the photon energy $\hbar\omega_{1}$, as shown in
Fig.~\ref{Fig-Pr141-Bi209-Aniso-Polari}, although this dependence is slightly reduced when compared to the case of
$^{71}$Ga$^{29+}$. In practice, this dependence arises from the interplay of the lifetime and splitting of the hyperfine levels
that contribute to the $\gamma_2$ emission, and it becomes strongest when the splitting and linewidths are comparable in
magnitude. For both $^{141}$Pr$^{57+}$ and $^{209}$Bi$^{81+}$ ions, moreover, the corresponding sensitivity coefficients are also
given in Table~\ref{Table-1}. The obtained strong angular and polarization dependence of the emitted $\gamma_{2}$ photons on the
hyperfine constant is therefore expected to help determine the hyperfine splitting of highly charged helium-like ions.

\textit{Experimental feasibility.}---The proposed scheme is feasible with present-day experimental facilities, such as, heavy-ion
storage rings or electron-beam ion traps. The initial $1s2s~^{1}S_{0}$ level can be populated quite selectively via $K$-shell
ionization of lithium-like projectiles in relativistic collisions with gas target at the experimental storage ring
\cite{Fritzsche/JPB:2005, Rzadkiewicz/PRA:2006, Trotsenko/PRL:2010}. Alternatively, it can also be populated via the prompt
$2s2p~^{1}P_{1}$ $\rightarrow$ $1s2s~^{1}S_{0}$ decay following the resonant electron capture of hydrogen-like ions into the
$2s2p~^{1}P_{1}$ resonance \cite{Mokler/PRL:1990}. Due to the high selectivity on the production of the $1s2s~^{1}S_{0}$ level,
as demonstrated in these experiments, the influence of neighboring levels on its subsequent decay can be ignored.

The $1s2s~^{1}S_{0}, F_{i}$ level of helium-like ions is known to decay primarily into the $1s^2~^{1}S_{0}, F_{f}$ ground level
via two-photon (2E1) emission \cite{Volotka/PRA:2011}. Since we wish to study the effect of the $1s2p~^{3}P_{1}, F$ hyperfine
splitting upon the angular distribution and linear polarization of the emitted $\gamma_{2}$ photons, cf.
Fig.~\ref{Fig-schematic}, our aim is to make the $1s2s ~^{1}S_{0}, F_{i}$ $\rightarrow$ $1s2p~^{3}P_{1}, F$ transition strong
enough to compete with the 2E1 decay and thus to populate the $1s2p~^{3}P_{1}, F$ levels. For this aim, this transition is
supposed to be stimulated by the incident laser photons $\gamma_1$ with suitable intensity and tunable energy $\hbar \omega_1$.
For $^{71}$Ga$^{29+}$ ions, for example, we obtain the required minimum laser intensity $3.2 \times 10^{5}$~W/cm$^{2}$ by using
Eq.~(35) in Ref.~\cite{Ferro/PRA:2011}. For both $^{141}$Pr$^{57+}$ and $^{209}$Bi$^{81+}$ ions, in addition, the corresponding
minimum intensities are also determined in a similar way to be $3.0 \times 10^{9}$~W/cm$^{2}$ and $3.6 \times
10^{11}$~W/cm$^{2}$, respectively. For such low laser intensities, the resulting Stark effect on the energy levels of ions is
negligibly small and hence can be ignored. These intensities are easily accessible with present-day laser sources from
near-infrared to extreme-ultraviolet photon energy regions. Since the linewidth of laser radiations in this energy range is much
smaller than the natural linewidth of the levels involved, the incident stimulating laser photons $\gamma_{1}$ can be treated to
be monochromatic.

Finally, let us discuss the measurement of the angular distribution and linear polarization of the emitted $\gamma_{2}$
fluorescence photons. The angular distribution can be accurately measured by an array of highly efficient solid-state Ge($i$)
detectors placed at different angles, while the polarization can be determined by means of two-dimensional position-sensitive
x-ray detectors and Compton scattering technique \cite{Tashenov/PRL:2006-2011, Weber/PRL:2010}. Moreover, due to recent progress
in the channel-cut silicon crystal polarimetry technique the polarization purity of x-ray photons have been measured with an
unprecedented level of accuracy, $\sim10^{-10}$ \cite{Marx/PRL:2013}. Since the required emission flux of the $\gamma_{2}$
photons for achieving such a high accuracy is mainly restricted by the amount of the production of helium-like ions, we may thus
expect an experimental uncertainty of about $10^{-4}$ in measuring the anisotropy parameter $\beta$ and the polarization $P_{1}$.
This uncertainty allows to determine the hyperfine constant on a level of accuracy $7 \times 10^{-5}$ for $^{71}$Ga$^{29+}$ and
of $3 \times 10^{-4}$ for $^{209}$Bi$^{81+}$, which would be well below the level of the contributing QED effects.

In summary, the angular distribution and linear polarization of x-ray photons emitted in a two-step radiative decay of highly
charged helium-like ions have been studied with the aim to pursue a scheme for determining their hyperfine splitting. For the
particular process (\ref{particular-process}), it is found that the angular and polarization behaviors of the $\gamma_{2}$
photons depend strongly on the hyperfine splitting of the $1s2p~^{3}P_{1}, F$ levels. This dependence will allow a determination
on the hyperfine splitting of helium-like ions with an accuracy of about $10^{-4}$, together with the hyperfine structures of
hydrogen- and lithium-like ions which could serve as a probe of QED in strong electromagnetic field generated by heavy nuclei.

We are grateful to S.~Trotsenko and B.~Marx for very helpful discussions on the production of highly charged ion beam and
high-precision x-ray polarization measurement, respectively. This work has been supported by the BMBF (Grant No.~05P15SJFAA) and
the NSFC (Grant Nos. 11464042 and U1332206).

\end{document}